\documentclass[journal]{IEEEtran}
\usepackage[switch]{lineno}%for drafting purposes
\usepackage{color}
\usepackage{hyperref}
\usepackage{enumerate}
\usepackage{verbatim}
\usepackage{nicefrac}
\usepackage{subcaption}
\ifCLASSINFOpdf
  \usepackage[pdftex]{graphicx}
   \DeclareGraphicsExtensions{.pdf,.jpeg,.png,.jpg}
\else
\fi
\usepackage[cmex10]{amsmath}
\usepackage{mathabx}
\usepackage{url}
\begin{document}
%\linenumbers%for drafting purposes
\title{Performance of a Quintuple-GEM Based RICH Detector Prototype}

\author{Marie~Blatnik,
        Klaus~Dehmelt,
        Abhay~Deshpande,
        Dhruv~Dixit,
        Nils~Feege,
        Thomas~K.~Hemmick,
        Benji~Lewis,
        Martin~L.~Purschke,~\IEEEmembership{Senior Member,~IEEE,}
        William~Roh,
        Fernando~Torales-Acosta,
        Thomas~Videb\ae k,
        and~Stephanie~Zajac% <-this % stops a space

%\thanks{This manuscript was submitted for peer review on January 13, 2015.}
\thanks{This work was supported in part by the U.S. Department of Energy (DOE) under award number 1901/59187.}
\thanks{M.~Blatnik, K.~Dehmelt, A.~Deshpande, D.~Dixit, N.~Feege, T.K.~Hemmick, B.~Lewis, W.~Roh, F.~Torales-Acosta,
        T.~Videb\ae k, and S.~Zajac are with the Department of Physics and Astronomy, Stony Brook University, Stony Brook,
NY, 11794-3800 USA (e-mail: klaus.dehmelt@stonybrook.edu).}% <-this % stops a space
\thanks{M.L.~Purschke is with Brookhaven National Lab, Upton, NY, 11973-5000 USA.}
%\thanks{Manuscript received January, 2015; revised ?}
}

\maketitle

\begin{abstract}
Cerenkov technology is often the optimal choice for particle identification in high energy particle collision applications. Typically, the most challenging regime is at high pseudorapidity (forward) where particle identification must perform well at high laboratory momenta. For the upcoming Electron Ion Collider (EIC), the physics goals require hadron ($\pi$, K, p) identification up to $\sim$~50 GeV/c. In this region Cerenkov Ring-Imaging (RICH) is the most viable solution.\newline
The speed of light in a radiator medium is inversely proportional to the refractive index. Hence, for particle identification (PID) reaching out to high momenta a small index of refraction is required. Unfortunately, the lowest indices of refraction also result in the lowest light yield ($\frac{dN_\gamma}{dx} \propto \sin^2{\left(\theta_C \right)}$) driving up the radiator length and thereby the overall detector cost. In this paper we report on a successful test of a compact RICH detector (1 meter radiator) capable of delivering in excess of 10 photoelectrons per ring with a low index radiator gas ($CF_4$). The detector concept is a natural extension of the PHENIX Hadron-Blind Detector (HBD) achieved by adding focusing capability at low wavelength and adequate gain for high efficiency detection of single-electron induced avalanches. Our results indicate that this technology is indeed a viable choice in the forward direction of the EIC. The setup and results are described within.
\end{abstract}


% \fi
\IEEEpeerreviewmaketitle

%----------------------------------------
%----------------------------------------
\section{Introduction}
\IEEEPARstart{T}he Electron Ion Collider (EIC) \cite{ref:eicwhitepaper}, envisioned to be constructed during the early 2020s will, for the first time, precisely image the gluons and sea quarks in the proton and nuclei. It will accelerate polarized electrons and a variety of light and heavy ions, from $H$ ($p$ and $d$) to $U$, of which the lightest ions can also be polarized. The EIC aims to completely resolve the internal structure of the proton and explore a new QCD frontier of ultra-dense gluon fields in nuclei at high energy. It has been given highest priority of the U.S. QCD community \cite{ref:townmeeting} for new construction.\newline
A detector capable of measuring the fragments of Electron-Ion collisions has to overcome a number of challenges. The particle flow is highly boosted into the forward direction due to the beam kinematics, which creates a very high density of high momentum particle tracks in the laboratory frame. A solution for Hadron particle identification (PID) is a Ring Imaging Cerenkov (RICH) detector with at least two radiators, with larger refraction index for covering small(er) momenta and small refraction index for covering the highest momentum range. The latter is usually achieved by using gas as radiator medium. However, the number of photoelectrons is limited due to the relatively small number of single photons ``produced" in the dilute medium.\newline
Our RICH concept has been developed with the goal of achieving PID at 50 GeV/c using only 1 meter of radiator gas but with sufficient number of photons. Our detector prototype was tested as
\begin{enumerate}
\item proof-of-principle test in an electron-test-beam environment at SLAC ESTB ({\href{http://estb.slac.stanford.edu}{{End Station (A) Test Beam}}}) and
\item under real-test conditions within a test-beam environment with various hadrons at various momenta at FTBF ({\href{http://www-ppd.fnal.gov/Ftbf/}{{Fermilab Test Beam Facility}}}).
\end{enumerate}

The windowless technology as well as the use of a wavelength-tuned mirror will minimize the loss of photons and will establish a successful operation of a RICH detector despite its short radiator length. Such detector will be extremely useful for experiments that require PID up to highest momenta but might be limited in space.\newline
In the following sections we describe the unique RICH technology, the detector prototype setup, and the operation at the test-beam facilities. Results from the tests will be described and discussed. 
%----------------------------------------
%----------------------------------------
\section{Windowless RICH technology}
For semi-inclusive and exclusive measurements in an EIC detector particle identification will be a critical element. One aims for hadron identification with better than 90\% efficiency and better than 95\% purity. Pions and kaons have to be positively identified, whereas for protons a negative particle identification suffices, {\it i.e.}, a particle is neither identified as an electron, pion, nor kaon. For higher momentum particles (forward direction) one needs to achieve these measurements to momenta of up to 50 GeV/c. Only RICH detectors with a low refractive index gas as radiator can operate in this regime. Tetrafluoromethane CF$_4$ has the lowest index of refraction among all fluorocarbons: $n_r(\lambda)-1\sim 5.5\times10^{\text{-}4}$ for 140 nm \cite{ref:cf4index}.\newline
Ring imaging requires the focusing of Cerenkov photons into a ring, whose radius is directly related to the Cerenkov angle. One option is to perform the focusing with a mirror and place the photo-detector at its focal plane.\newline 
Cerenkov light yield follows the relation \cite{ref:cherenkovtamm1,ref:cherenkovtamm2}
\begin{equation}
\frac{dN_{p.e.}}{dx}=2\pi\alpha Z^2 \sin^2\theta_C\int_{\lambda_{min}}^\infty\varepsilon(\lambda)\frac{d\lambda}{\lambda^2}~.
\label{eq:numberphotons}
\end{equation}
Here $\varepsilon(\lambda)$ is an overall efficiency which leads to the detection of the number of photo-electrons N$_{p.e.}$ from the number of photons N$_\gamma$ and includes effects due to absorption, reflection, transmission, and detection probability. $Z$ is the charge of the incident hadron in units of the elementary charge. The most effective way to maximize $\frac{dN_\gamma}{dx}$ is to pursue low wavelengths as driven by the $\frac{1}{\lambda^2}$ term from Eq.~\ref{eq:numberphotons}, and also by the tendency of most photosensitive materials to have increasing quantum efficiency with decreasing wavelength. Consequently, the detection of Vacuum Ultraviolet (VUV with 200 nm~\textgreater ~$\lambda$~\textgreater~100 nm) photons should be the principle goal in making the most compact detector design. Challenges are provided by the fact that common mirror materials have a cutoff wavelength above the VUV-range and that many common photo-detectors use cathodes that are physically separated from the radiator medium by a window. To overcome these challenges several technologies have been implemented in the presented prototype setup making it the first windowless RICH detector:
\begin{enumerate}[A.]
\item Windowless photocathode
\item VUV high reflective mirror coating
\item Quintuple GEM photo-detector
\end{enumerate}
%----------------------------------------
\subsection{Quintuple GEM}\label{sec:5gem}

\begin{figure}[!ht]
\centering
\includegraphics[width=3.25in]{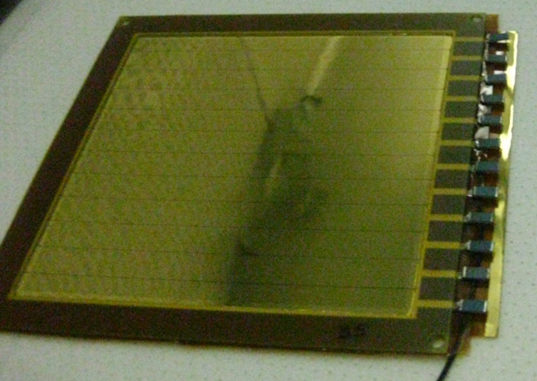}
\caption{(Color online) RICH Gem featuring 12 HV sectors, 20 M$\Omega$ resistors, and Cu-Ni-Au surface (produced at the CERN PCB workshop).}
\label{fig:segGEM}
\end{figure}

\begin{figure}[!ht]
\centering
%\vspace*{-0.5in}
\includegraphics[width=3.5in]{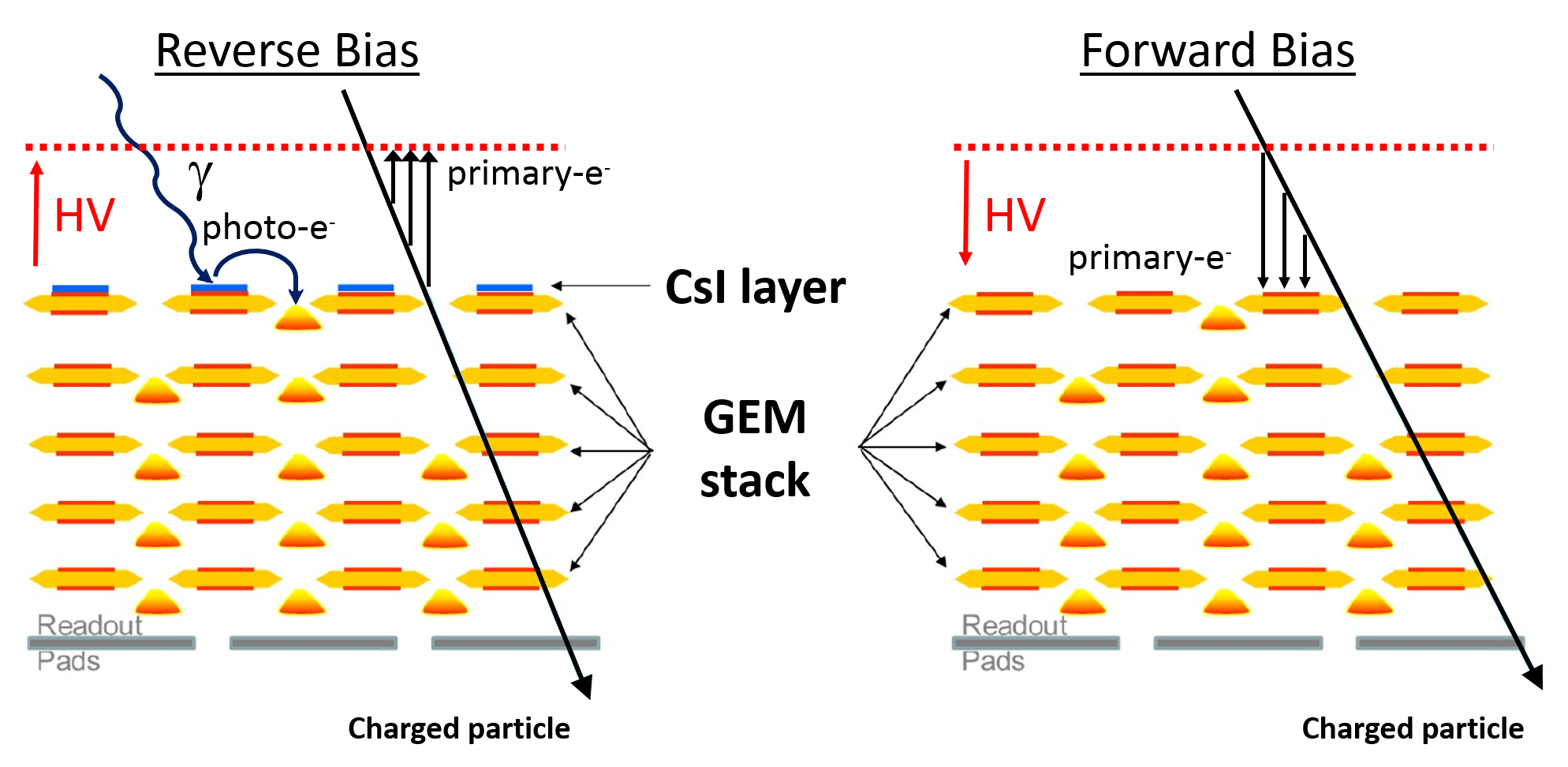}
%\vspace*{-0.5in}
\caption{(Color online) Quintuple-GEM configuration with readout pads anode. Five GEMs with a distance of $\sim 1.6$ mm from each other serve as amplification device with high efficiency for single photoelectrons. The top GEM is covered with a CsI layer providing a photoelectrode. Left: reverse bias mode for rejecting primary ionization electrons. Right: Forward bias mode for use as track position sensitive device (without photoelectrode). }
\label{fig:quintuplegem}
\end{figure}

The Gas Electron Multiplier, invented by F.~Sauli in the mid-1990s \cite{ref:sauligem} serves as the general amplification structure for the RICH detector prototype. A thin polymer foil with copper-cladding, perforated with a high density of microscopic holes 
%(see Fig.~\ref{fig:gemsem}),%
acts as the element for electron multiplication due to applied high-voltage across the capacitor-like structure. The foils can be stacked such that the amplification load is reduced at each element, allowing for safe operating conditions. For using GEMs as photomultipliers, two steps are required. First the GEM surface itself should be made non-reactive with the photocathode material. As such the GEMs used for our research have Ni and Au over-coatings to hide the copper (see Fig.~\ref{fig:segGEM}). Second we deposit a thin layer ($\sim$~300 nm) of CsI on top of the GEM that is facing the radiator medium (see Fig.~\ref{fig:quintuplegem}). CsI is widely used in gaseous photon detectors as well as in RICH applications, for example in the HBD in the PHENIX \cite{ref:phenix} experiment. The  procedure  for  covering  the GEMs  with CsI,  which we followed is described in detail in \cite{ref:hbd4phenix}. This research is correctly characterized as an evolution of the HBD effort to a focused design. More details are discussed in Sec.~\ref{prototypesetup}.\newline
Based on the experience with the HBD in PHENIX we do not expect aging effects, also, because the overall rate in a future EIC detector is low compared to Heavy Ion experiments where the HBD was situated. This is also expected for the LHCb experiment \cite{ref:lhcbaging}.
\newline

The motivation to use as many as five GEMs stems from two factors. First, we must achieve high detection efficiency for avalanches induced by single photo-electrons. Experience with the PHENIX HBD showed that the application of the CsI layer could, in some cases, limit the gain capability of that particular GEM. The PHENIX HBD (triple GEM) was run at a gain of 4000, which resulted in the hardware zero-suppression being set at a pulse height equal to the mean signal generated by a single photoelectron avalanche. The gain process will yield a Polya distribution (well approximated here by an exponential) indicating that the PHENIX efficiency for single photoelectron avalanche detection was $\frac{1}{e} \sim 37\%$, unacceptably low for our current application. Although it can well be expected that a fourth GEM would provide more than sufficient additional gain (see, {\it e.g.} \cite{ref:4gemgain}), a secondary concern arose during the design phase. The success in achieving high gain from a photosensitive GEM stack is partly due to the fact that the photocathode itself is optically shielded from the avalanche. By introducing a mirror into our RICH, we risked background light from avalanches in the first hole layer ($CF_4$ produces light at $\sim$~160 nm during avalanche) being reflected back onto the photocathode surface. The addition of a fifth GEM allowed us to optionally operate the photosensitive GEM at a gain near 1 (using ~80\% of the nominal $\Delta V$) to minimize this possibility. During our SLAC run we verified that:
\begin{enumerate}
\item We could indeed operate the detector at gain=1 in the top layer with no measurable signal loss.
\item Any photon feedback from the top layer reflected by the mirror was negligible.
\end{enumerate}
Thus, the eventual application of this technology is likely feasible using only four GEMs. As a final precaution against optical feedback, each GEM was stacked with a 90 degree rotation as compared to its neighbors, thereby making a minimal and uniform optical transparency.\newline
Scintillation light is in general a potential problem for Cerenkov counters using highly UV transparent gases such as CF$_4$. However, for the detection of particles with velocities $\beta \rightarrow 1$ this will not pose a problem in measuring the signal due to the high granularity of the readout.\newline
A stack of five GEMs, to our knowledge, has not been used before. We routinely achieved gains in excess of 10$^5$ in the five GEM configuration without a single spark over an integrated time of roughly four weeks in beam. More details on this issue will be provided in Sec.~\ref{prototypesetup}.\newline
Because of the large number of amplification elements electrons from primary ionization in the gas volume of the transfer gaps will be amplified, yielding hits. In this sense, even when operated in reverse bias, this detector cannot be considered as truly "Hadron Blind". Nonetheless, by having the ionization trail of ionizing particles undergo one fewer avalanche step, the "MIP" hit becomes of merely comparable average pulse height to that induced by a single photoelectron. The relative size of photoelectron signals and ionization trails can be further tuned by optimizing the transfer gap sizes and the transmission efficiency of electrons from previous amplification elements.\newline
A resistor chain supplied a potential difference across the GEMs for amplification, transmission of electrons through the GEMs and the transfer/induction gaps. An independent power supply between a mesh and the top of the first GEM provided an independent potential difference. The mesh was made from stainless steel with 88\% optical transparency and allowed photons to pass through with high probability. Electrons liberated by the passage of a charged particle in the gas volume between the mesh and the top of GEM1 might avalanche and thereby mimic the signal from true photoelectrons. The so-called "reverse-bias" configuration is achieved by setting the mesh voltage less negative than the top of GEM 1. In reverse bias mode these ionization electrons will drift towards the mesh, thus preventing them from being amplified. So long as the reverse bias is very slight very few photoelectrons drift to the mesh. These electrons will predominantly experience the electric field from the capacitor-like structure and be directed into the hole. Furthermore, it is well known that operation of a photocathode in the presence of gas can result in electron loss due to collisions with the gas molecules that "reflect" the electrons back into the cathode. This effect is minimized by having a strong electric field at all points along the photocathode surface. The electric field characteristics of the "standard" GEM hole geometry (70 $\mu$m holes, 140 $\mu$m pitch) were analyzed in great detail for the PHENIX HBD and found to be a virtually ideal cathode which (when operated with a SMALL reverse bias) both efficiently produced and transported photo-electrons to the holes. Larger pitch devices ({\it e.g.} some thick GEM geometries) may be less efficient as photocathodes unless carefully tuned via detailed electrostatic calculations.\newline
The quintuple-GEM setup has been tested regarding its gain value and uniformity. A radioactive source ($^{55}$Fe) with well known intensity was used to create a signal for the quintuple-GEM. The procedure will be further discussed in Sec.\ref{sec:preparation}. A known amount of charge is deposited in the drift volume of the GEM-detector, in forward bias, and compared with the charge reaching the readout and processed electronically based on the APV25 chip \cite{ref:apv25} and a Scalable Readout System (SRS \cite{ref:srs}) as back-end. Varying the potential across the GEM-stack results in various gain settings.\newline
The readout plane of the detector is a square array composed of 512 tessellated hexagons with an apothem of $\sim$2.5 mm to detect and determine the position of the individual photoelectrons.\newline
The APV25 chip has 128 readout channels, each consisting of a 50 nanosecond CR-RC type shaping amplifier, a 192 element deep pipeline and a pulse shape processing stage.\newline
The SRS is designed around a bivalent scalability concept and introduces a modular concept that offers the possibility to connect different front-end ASICs to the standard SRS electronics, allowing the user to choose the most suitable front-end for the detector technology employed. For the presented setup four APV25 chip cards have been connected to the readout pads with two cards each daisy-chained and connected to the FECs.\newline
The noise introduced in the described setup was identified to be $\sim$~600 e$^-$. For single photon detection with a S/N of 10/1 a gain of less than 10$^4$ is required.\newline
For obtaining a reliable signal one has to trigger the readout of the signal outside the noise-band. Dependent upon the reliability requirement of the signal readout one chooses typically several sigmas of the noise-level. With increasing sigma-levels the efficiency of collecting events decreases while increasing the purity of events, {\it i.e.}, the chance of collecting real events and not noise.\newline
In order to avoid damages due to electric breakdowns one has to minimize the energy that can be stored on the GEM. The stored energy in a capacitor is $E=\frac{1}{2}CV^2$, with $C$ the capacitance and $V$ the applied voltage. Consequently, one can either reduce the capacitance or voltage applied across the capacitor, or both. The reduction in capacitance can be achieved by reducing the area of the capacitor. For a GEM foil one can subdivide the surface into smaller areas, {\it i.e.}, introducing electrically insulated sectors on the surface. Our GEMs were divided into 12 sectors across the 10$\times$10 cm$^2$ surface as shown in Fig.~\ref{fig:segGEM}. This high segmentation is driven by the fact that using CF$_4$ as the avalanche gas moves the operating voltage across each GEM to nearly double that used in more common gases ({\it e.g.} Ar:CO$_2$ 70:30). High voltage (HV) across the multi-GEM stack was provided through a resistor chain.% (Fig.~\ref{fig:resistorchain}). 
%----------------------------------------
\subsection{VUV mirror}\label{sec:mirror}

Due to the limited reflectivity of typical regular mirror coatings for photons at smaller wavelengths a specialized mirror technology has been deployed by commercial partners that provides sufficient reflectivity deep in the VUV. This technology uses a carefully tuned thickness of MgF$_2$. The focusing mirror is polished (RMS=20 \AA) fused silica coated with Al and MgF$_2$ approximately 250 \AA~thick, so that thin film effects are playing a role: MgF$_2$, under regular circumstances is not transmissive below $\sim$~140 nm. At this thickness the overcoating not only protects the Al from oxidation, but also serves as a thin-film reflector with a peak reflectivity at $\lambda$=120 nm. The relatively small mirror ($\diameter$ 18 cm with curvature radius of 2 m) for the detector prototype was purchased from Acton ({\href{http://www.princetoninstruments.com/products/optics/mirrors/}{{Princeton Instruments, Acton Optics \& Coating}}}). Due to prohibitively high cost, we will attempt to develop a large mirror ourselves for deployment at the EIC.\newline
%\IEEEpubidadjcol
%----------------------------------------
%----------------------------------------
\section{RICH Detector Prototype}
\subsection{Detector Prototype Setup}\label{prototypesetup}
\begin{figure}[!ht]
\centering
\vspace*{-0.5in}
\includegraphics[width=3.25in]{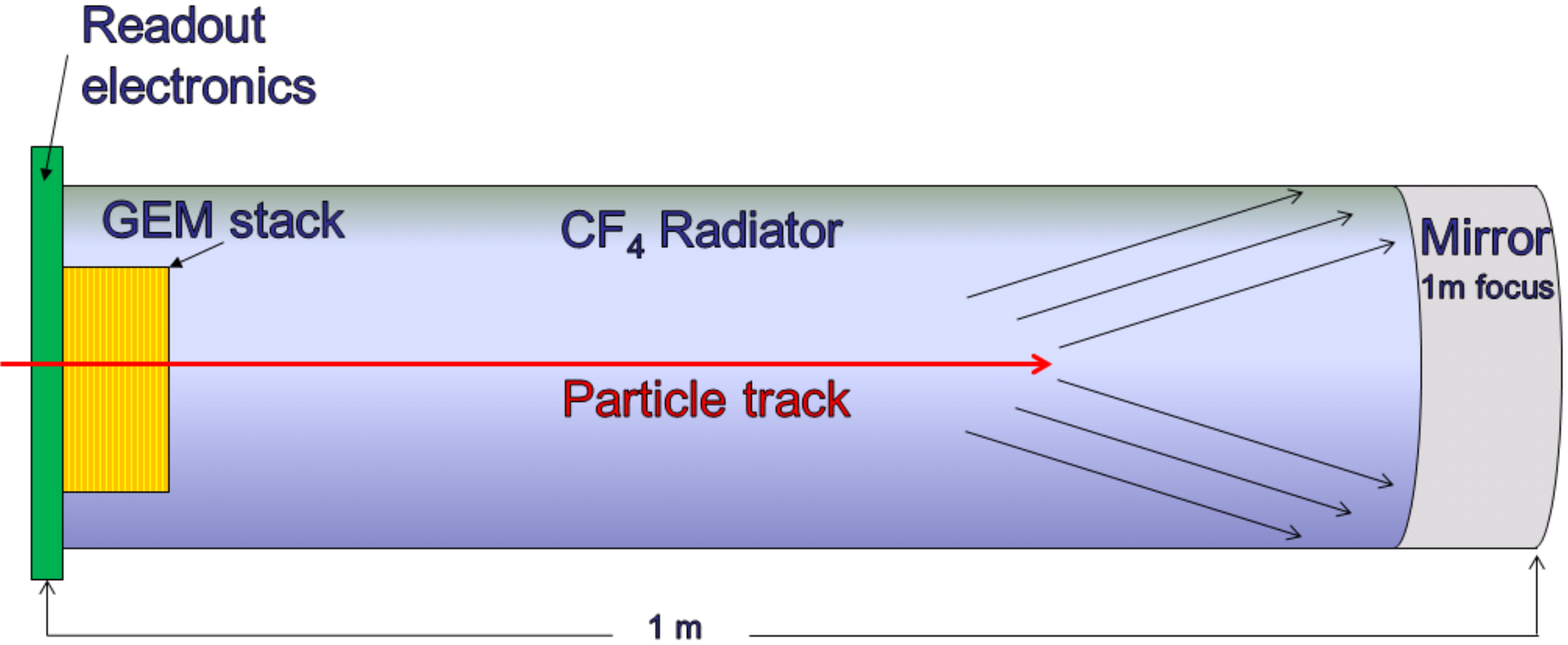}
\vspace*{-0.5in}
\caption{(Color online) Schematic diagram of the RICH detector prototype setup.}
\label{fig:detectordiagram}
\vspace*{5mm}
\includegraphics[width=3.25in]{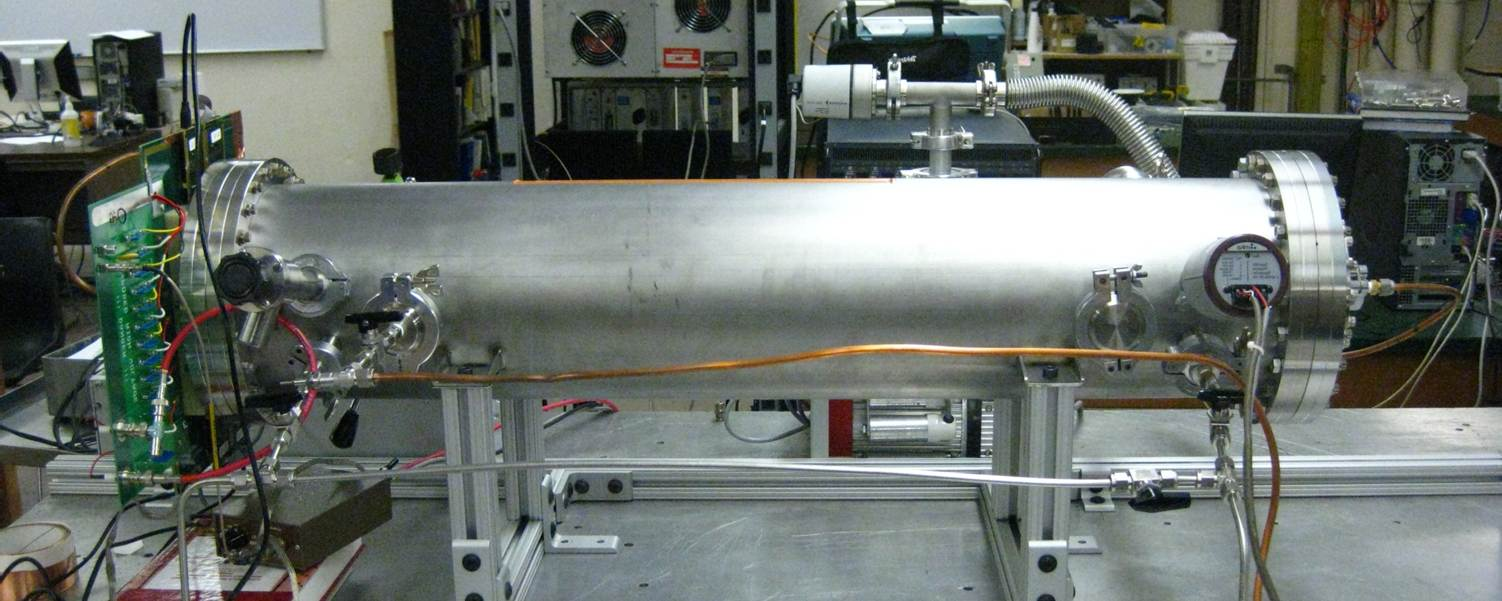}
\caption{(Color online) Photograph of the RICH detector prototype setup.}
\label{fig:detectorphoto}
\end{figure}

The detector prototype consists of a one meter long, 20 cm radius cylindrical tank made out of stainless steel. The cylinder was designed such that one can apply high vacuum ($10^{-6}~Torr$) conditions to the setup and to allow for various insertions  ({\it e.g.} $^{55}Fe$ for testing) and quick gas exchange without breaking the seal. Neither the detector plane nor the mirror could mechanically withstand a differential pressure of one atmosphere. Vacuum conditions were therefore applied to both sides of these devices by installing massive removable covers (flanges) over their exterior surfaces. These covers were removed while accepting beam.\newline
The quintuple-GEM module acted both as the detection plane and as the entrance window. The mirror, located at the far end of the tube, also served double-duty by being both the focusing device and the exit window. A diagram of the setup can be seen in Fig.~\ref{fig:detectordiagram} and a photograph of the setup in Fig.~\ref{fig:detectorphoto}. Standard 10$\times$10 cm$^2$ GEMs with a hole diameter of 70 $\mu$m and 140 $\mu$m pitch were used.\newline
In a test-beam environment the charged particle traverses the radiator, from left to right (Fig.~\ref{fig:detectordiagram}). The photons of the Cerenkov cone are reflected from the mirror of the right side and focused onto the readout plane on the left side. The curvature radius of the mirror is 2 m such that the focal plane of the mirror coincides with the surface of the CsI-coated GEM. Here, photoconversion takes place and the photoelectrons released are accelerated into the GEM structures and undergo multiplication to produce a sizable electronic signal on the readout plane.\newline
CsI-coating is a well established procedure and was successfully used for the HBD \cite{ref:hbd} for the PHENIX experiment.
So long as the photocathode exceeds a certain minimum thickness of photosensitive material (enough to absorb the photon) the quantum efficiency is high. This minimum thickness was measured to be $\sim$~200 nm for CsI \cite{ref:csicathode}, and the described detector prototype used a layer of 300 nm or more. A reproducible high quantum efficiency of up to 70\% at smallest wavelengths ($\sim$~120 nm, see also \cite{ref:csicathode}) has been achieved which allows a very efficient use of CsI-GEMs as photocathode material.\newline
A recirculating gas system, providing the CF$_4$ radiator with high purity (grade 5.7), was constructed and used especially for the test-beam setup. Recirculation, for saving the rather costly radiator gas, requires a purification system that removes those contaminants which negatively affect the operation of the RICH. In particular, H$_2$O can cause the CsI-layer on the GEM foil to change its composition and degrade the quantum efficiency. O$_2$ is electronegative and will reduce the electron avalanche needed for a proper signal readout, and more importantly, reduces the transparency in the gas for VUV photons as does H$_2$O. These contaminants were reduced with commercial purifiers (Cu-catalyst and molecular-sieves \cite{ref:purifiers}) well below the 1-ppm-level.\newline
The gas system was equipped with a pump, mass flow controllers and a manometric system so that it could provide the detector system with a constant, slightly pressurized radiator medium ($\sim$~2~Torr above ambient).
\subsection{Preparation}\label{sec:preparation}
The setup was tested in its final configuration so that it was ready to be operated in a test-beam environment. The preparational tests consisted of gain calibration, pedestal correction, and common-mode noise calibration. These tests were performed before the setup moved to one of the test-beam sites and were also partially performed while at the test-beam site for verifying the functionality of the detector prototype.\newline
\subsubsection*{Gain Calibration}
\begin{figure}[!ht]
\centering
%\vspace*{-0.25in}
%\hspace*{-0.15in}
\includegraphics[width=3.8in]{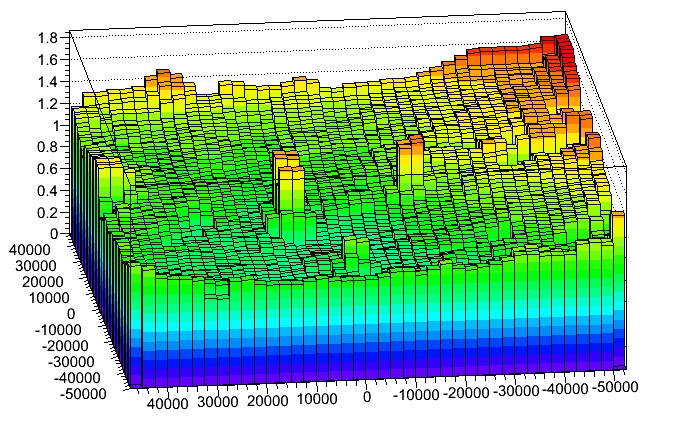}%pdf}
%\vspace*{-0.5in}
\caption{(Color online) Variation of gain across the area of the GEM amplification module, measured across the pad plane at a medium stable gain setting. The {\textit x/y}-axis show the dimensions along the GEM area in microns. The vertical axis shows the signal strength and varies slightly at few places, but in particular in one corner of the GEM significantly. We suspect that a non-uniform separation between the GEM layers along the mounting posts might be responsible for this behavior.}
\label{fig:gainfactor}
\end{figure}

Calibration of the gain was accomplished by using an $^{55}$Fe source which could be inserted through a flange with a retractable arm. This allowed the source to be moved across the surface of the GEM.\newline
The source produced on average 110 electrons from each X-ray. The resulting signal throughout the GEM-stack was read-out via a capacitive pick-off circuitry which was attached to the bottom surface of the last GEM along the stack. Fig.~\ref{fig:gainfactor} shows the gain map across the whole area of the GEM module. The variation might be explained by a non-uniformity in flatness of GEMs throughout the stack. As can be seen from the colored surface, most of the area has variations of $<\sim 10$\% and thus well within tolerable levels.\newline
The efficiency of picking up the signal above threshold depends on the gain setting and the threshold cut applied. If we apply a threshold cut of 3$\sigma$ we obtain at a gain of about 4.5$\times10^4$ an efficiency of better than 98\%, and when we increase the cut to 10$\sigma$ we obtain at the same gain an efficiency of slightly less than 90\%.
%----------------------------------------
%\newpage
\section{Test-Beam SLAC-ESTB}
\subsection{Setup}
The detector prototype setup was driven by truck to SLAC and set up in the beam line of SLAC ESTB. The vessel was sealed and a low flow of dry gas (several turns per day) was maintained during the trip from New York to California. The main task was to test the functionality of the detector prototype with single electrons. Electrons saturate the Cerenkov angle at already very low energies, hence no velocity measurement of the electrons was necessary.\newline
For the test-beam operation of the RICH detector prototype a 5 Hz, 9 GeV electron beam was collimated into the test-beam area. With 70\% probability the electron bursts were empty (no electrons). That means that to a large extent bursts with two or more electrons were excluded. For triggering and selecting in the analysis stage a plastic scintillator and Lead-Glass (PbGl) calorimeter were placed downstream and included into the data acquisition via a DRS4 chip-based readout system \cite{ref:drs4}. Signals from these devices were used to identify and reject spills with other than one electron incident on the RICH.
\subsection{Results}
\begin{figure}[!ht]
\centering
%\vspace*{-0.25in}
\includegraphics[width=3.5in]{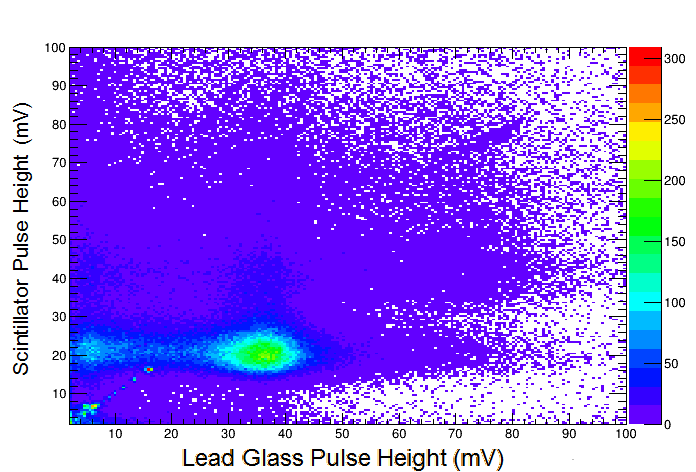}%pdf}
%\vspace*{-0.5in}
\caption{(Color online) Correlation of pulse heights of scintillator and PbGl calorimeter. The prominent region corresponds to pulse heights that originate from single electrons.}
\label{fig:scintcal}
\end{figure}

Data were collected for about 80 hours of beam-time. Each run consisted of roughly 30 minutes. Events were triggered by the beam clock, i.~e., all events during the spill were collected. This allowed for about 9000 events to be recorded for each run. Events that contained single electrons were filtered by using scintillator and calorimeter information. This was obtained by selecting events according to a correlation between scintillator pulse height and calorimeter pulse height; see Fig.~\ref{fig:scintcal}.\newline
Data were analyzed according to the identification and the measurement of the ring diameters. A pattern recognition algorithm was used to identify the pads that were producing the ring and a fitting procedure was used to determine the position and diameter of the ring \{x,y,d\}.\newline

\begin{figure}[!ht]
\centering
\includegraphics[width=2.5in]{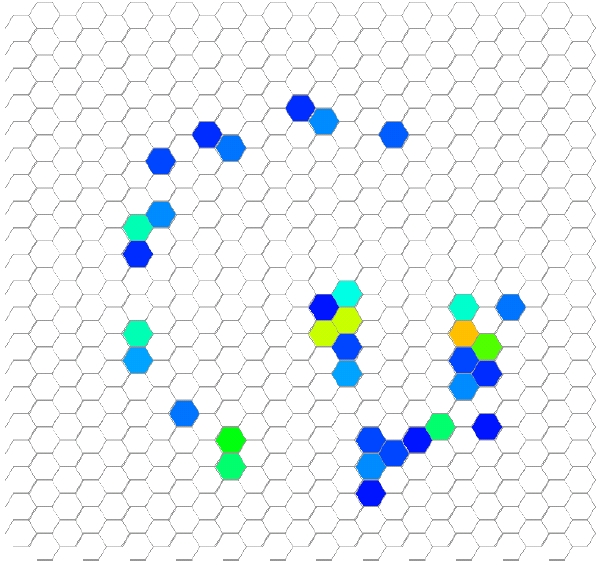}
\caption{(Color online) One of the first Cerenkov rings recorded at the ESTB. Clearly visible is the shape of the ring and because the GEM cascade was operated in forward bias the minimum ionizing part of the electron can be seen in the center of the ring. This single ring shows "gaps" which are purely due to photon statistics.}
\label{fig:firstring}
\end{figure}
\subsubsection{Combinatorial Hough Transform}
A combinatorial Hough transform (CHT \cite{ref:cht}) algorithm has been applied to identify all possible hit combinations that make up a ring. The ring parameters were determined by choosing the most probable combination of the \{x,y,d\}-triplet. A  typical example of a ring that is taken as input for CHT can be seen in Fig.~\ref{fig:firstring}.\newline
%Background hits that occurred were eliminated as described in Sec.~\ref{sec:ftbf_results}.
Hits outside the expected ring region were identified as background. This was confirmed with the test-beam campaign at Fermilab in conjunction with tracker data, where shadow track hits appeared. It was found that cross-talk between neighboring channels in the APV readout was responsible for this background.\newline
Background was determined by a procedure that subdivided the readout board into four quadrants so that each APV25 card could be considered individually. In each column of a single quadrant a signal was accepted, {\it i.e.}, not to be background if it registered with a charge above the pedestal value plus a predetermined width ($\sigma$) and if it was the hit with the largest value in its half-column of pads. This procedure guaranteed that only a maximum of one hit was present within a single column of each quadrant. This algorithm apparently lowers the count of pads for a ring compared to the photon count. A Monte Carlo study was performed to estimate a possible degradation because of this rejection cut.\newline
%%%%%%%%%%%%%%%%%%%% BEGIN COMMENTED Area %%%%%%%%%%%%%%%%%%%%
%%%%%%%%%%%%%%%%%%%% End COMMENTED Area %%%%%%%%%%%%%%%%%%%%
\subsubsection{Charge distribution}
Charge deposits on readout pads, the number of Cerenkov-photons produced, and the number of photoelectrons released have been investigated.
\begin{figure}[!t]
\begin{subfigure}{0.5\textwidth}
\centering
%\hspace*{-0.85in}
\includegraphics[width=3.5in]{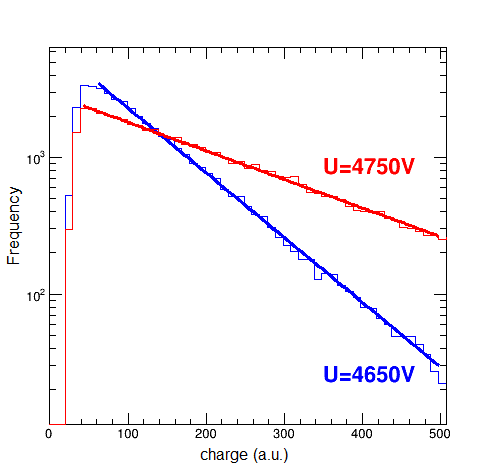}%df}
\end{subfigure}
\begin{subfigure}{0.5\textwidth}
\centering
%\vspace*{-0.3in}
\includegraphics[width=3.5in]{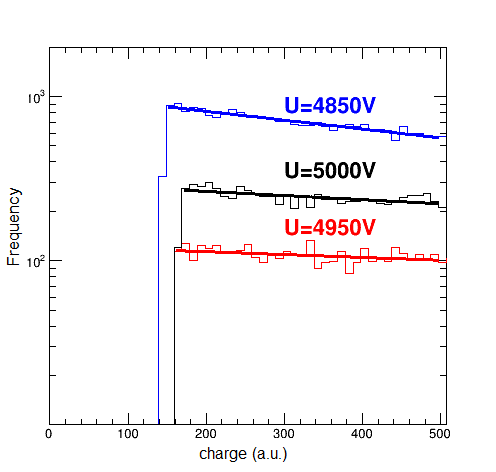}%df}
\end{subfigure}
\caption{(Color online) Charge spectrum for various gain settings. Upper: two rather low gain settings show the exponential spectrum for deposited charge on pads and its increasing steepness of the slope. Lower: for higher gain settings saturation and firing of neighboring pads resulted which caused a low pulse height component to appear which was cut away.}
\label{fig:qspec}
\end{figure}
A charge spectrum has been obtained according to Fig.~\ref{fig:qspec}: the upper graph represents the classic pulse height distribution behavior according to an exponential form at lower gains, and the lower graph represents a Polya distribution at higher gains \cite{ref:sergemmigas}.\newline
The number of responding pads saturates at large gain at a value around 9. The number of responding pads did not necessarily correspond to the number of photons that were contributing to the signals on the pads. For instance, one photon could lead to a response on two pads, or two photons could lead to a response on one pad only. A Monte Carlo simulation was performed to compare with the measured number of photons. It took into account the transverse diffusion of the charge cloud during the amplification process, the wavelength dependence of the refractive index of CF$_4$, resulting in a Cerenkov-angle as a function of wavelength, as well as a weighting effect for the Cerenkov intensity according to Eq.~\ref{eq:numberphotons}.
\begin{figure}[!ht]
\centering
%\hspace*{-.2in}
%\vspace*{-0.2in}
\includegraphics[width=3.75in]{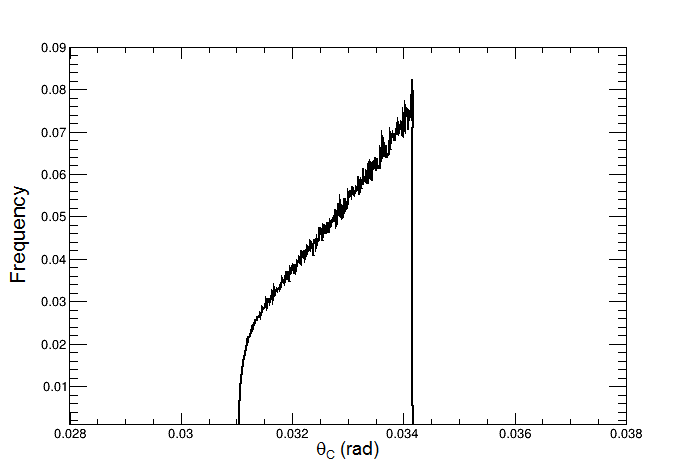}%df}
%\vspace*{-0.2in}
\caption{(Chromatic dispersion for the Cerenkov angle weighted according to Cerenkov intensity and quantum efficiency of a CsI cathode.}
\label{fig:dispdia}
\end{figure}
The angular dispersion of the Cerenkov angle as a result of chromatic dispersion was found to be $\sigma_{\theta_C}/\theta_C\sim$~2.5\% (Fig.~\ref{fig:dispdia}).
Furthermore, it was found that on average 9 photons were contributing to the Cerenkov ring. This is in discrepancy to the expected number of photons, which was 16 based on Eq.~\ref{eq:numberphotons} and an overall quantum efficiency of about 65\%, including photon absorption by the medium, reflectivity of the mirror and quantum efficiency of the photocathode. It is under investigation where this discrepancy is resulting from; it is possible that the quantum efficiency of the photocathode might have degraded during transport. The transport lasted four days across the country under relatively low gas flow. For the Fermilab tests, the gas flow was dramatically increased (few turns per hour instead of per day) and indeed, the photo-electron yield increased (see Sec.~\ref{sec:ftbf}).\newline
The size of the charge cloud along the amplification path can be calculated with a diffusion constant that has been determined  with simulations programs GARFIELD \cite{ref:garfield} and MAGBOLTZ \cite{ref:magboltz}. An extensive compilation for various gas compounds can be found at {\href{http://www-hep.phys.saga-u.ac.jp/ILC-TPC/gas/}{\textit{Saga University: ILC-TPC Gas Properties}}. The transverse diffusion in CF$_4$ through five transfer gaps of each 1.6 mm yields $\sigma_{transv}=60~\mu$m. However, an ``additional" diffusion was accomplished because of the misalignment of the hole pattern over the five GEM foils. This was calculated to be $\sigma_{hole}=118~\mu$m with a Monte Carlo method. Adding in quadrature one obtains a total transverse size of the charge cloud of $\sigma_{total}=132~\mu$m.
\begin{figure}[!hb]
\centering
\vspace*{-0.175in}
\includegraphics[width=3.75in]{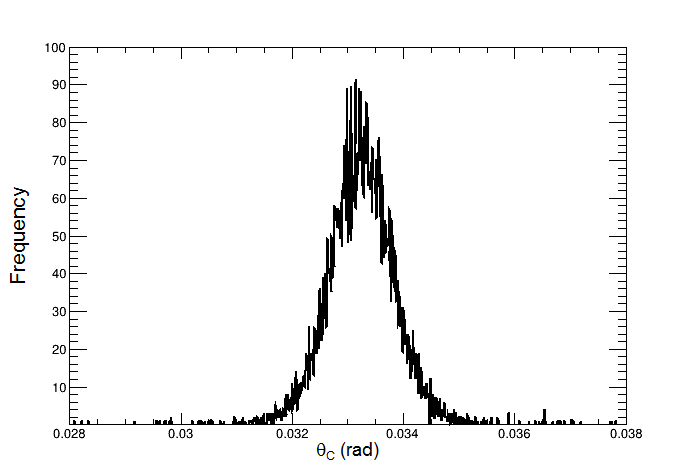}
\vspace*{-0.2in}
\caption{Simulated Cerenkov angle for pions with momentum of 32 GeV/c in a CF$_4$ radiator. A Gaussian fit yields $\theta_C=(33.2\pm 1.0)~mrad$.}
\label{fig:radiussim}
\end{figure}
Fig.~\ref{fig:radiussim} shows the result of the simulation and the Cerenkov angle to be 33.2 mrad with $\sigma_{\theta_C}/\theta_C=$ 3.0\%. The measured Cerenkov angle and width is in excellent agreement with the simulation.
%----------------------------------------
\section{Test-Beam FTBF}\label{sec:ftbf}
\subsection{Setup}
The main goal of the FTBF testbeam was to assess the detector performance for the identification of hadrons. Secondary particles were produced by 120 GeV/c proton beam indecent upon a target. The momentum range of the secondaries is 1-32 GeV/c, limited by the beamline optics. Although higher momentum secondaries are available from the "upstream target" most kaons produced upstream would have decayed before reaching our apparatus, limiting our experiment to the downstream target and a momentum of 32 GeV/c.\newline
The same setup as used at ESTB was used at FTBF. The CsI-coated GEM foil that has been used at SLAC was thoroughly washed off its CsI with de-ionized water and dried with alcohol. After this procedure a fresh CsI coating was applied to the same GEM.\newline
During the SLAC test, the rings were all intense enough that they could be used to self-determine their own centers. At Fermilab, some of the mass/momentum combinations were close enough to the Cerenkov threshold that one expects a rather low photon yield and these rings could not self-determine their own centers. To overcome this limitation, two $10\times 10~cm^2$ triple-GEM tracking detectors were used.\newline
For discriminating particle species a differential Cerenkov counter provided by FTBF was used; see Fig.~\ref{fig:diffcherenkov}. Two photo-multipliers (PMT), labeled as ``inner" and ``outer" PMT can selectively be used, dependent on the ring diameter of the particle under consideration. The mirror M2 has a small hole in its center which allows small ring radii to pass onto inner PMT, else will be reflected to outer PMT and counted.
\begin{figure}[!ht]
\centering
%\vspace*{-0.75in}
\includegraphics[width=3.5in]{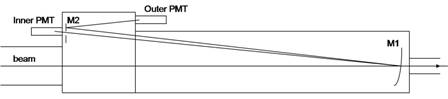}%pdf}
%\vspace*{-0.75in}
\caption{(Schematics of the FTBF differential Cerenkov counter.}
\label{fig:diffcherenkov}
\end{figure}
The pressure of the radiator gas could be tuned so that with varying $n_r$ the threshold and ring diameter of the Cerenkov signals velocities of the different massive particles could be distinguished by selecting signals only from the inner, respectively outer PMT. The consequence of this setup is that kaons would give a signal in the inner PMT, pions in the outer PMT, protons in neither.\newline
The downstream target only provided $\sim$~1\% content of kaons. For this reason, a trigger on the inner Cerenkov counter to take ``kaon-only" data was developed. The running period was divided such that 50\% of the clock time was dedicated to kaon trigger data.\newline
The Cerenkov threshold for kaons in CF$_4$ is $\sim$~16 GeV/c. FTBF was optimized for tunes in 5 GeV/c increments which allowed for three beam momenta useful for collecting data: 20 GeV/c, 25 GeV/c, and 32 GeV/c.
\subsection{Results}\label{sec:ftbf_results}
For intense rings, well above the Cerenkov threshold, it is possible to determine the ring center using the ring itself. The tracker/mirror alignment was fine-tuned by plotting the correspondence between the self-determined ring center with the direction vector of the tracks from the trackers. Once the system was aligned (using only intense rings), rings of any intensity were found solely based upon the tracker-determined ring center. 
\begin{figure}[!ht]
\begin{subfigure}{0.5\textwidth}
\centering
\includegraphics[width=3.5in]{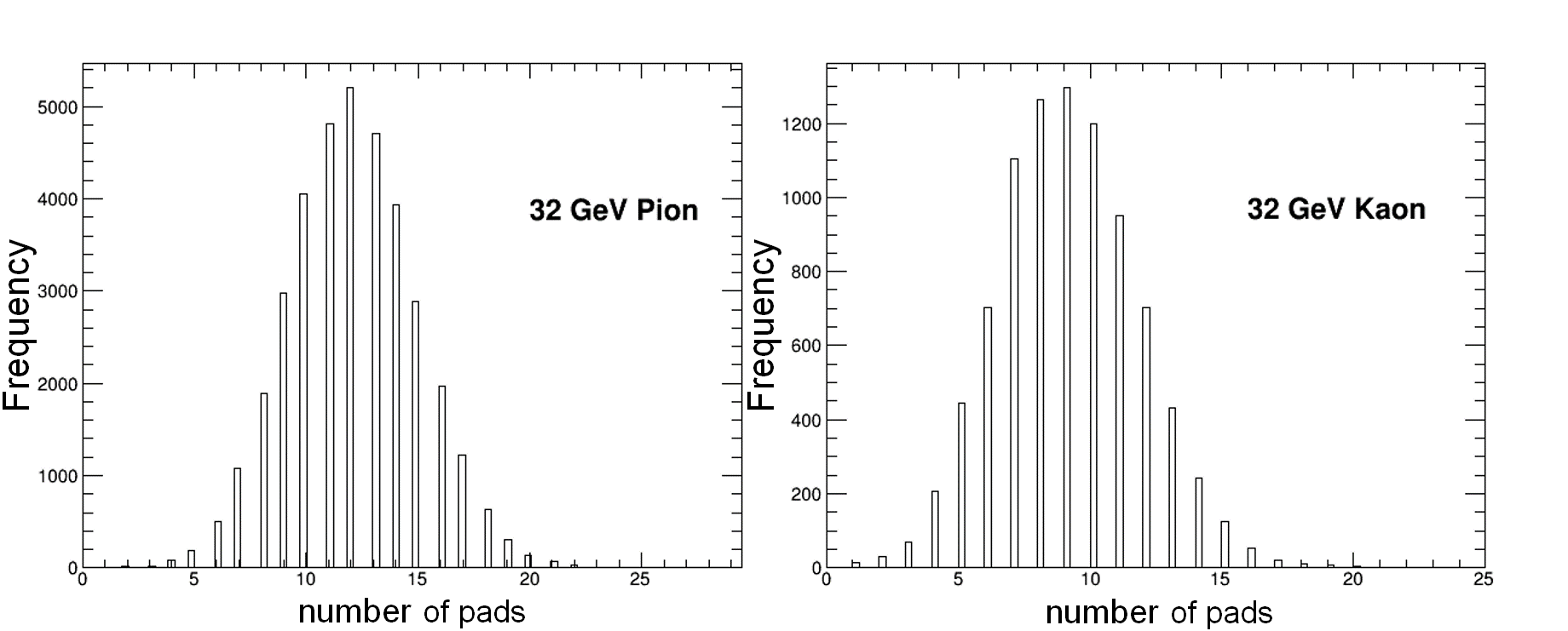}%pdf}
\end{subfigure}
\begin{subfigure}{0.5\textwidth}
\centering
\includegraphics[width=3.5in]{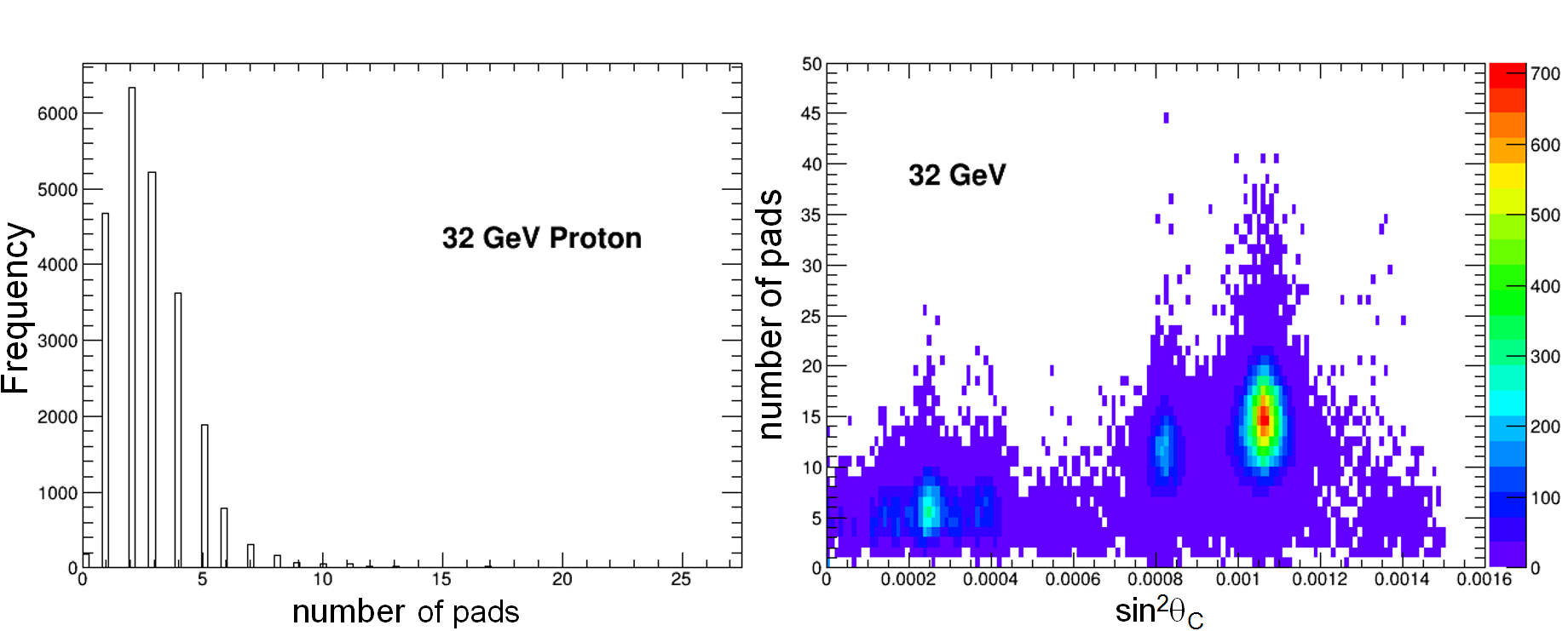}%pdf}
%\vspace*{-0.3in}
\end{subfigure}
\caption{(Color online) Distributions for the number of responding pads for various particles. Lower right: dependence of the $\sin\theta$ squared, which corresponds to the squared ring diameter, on number of responding pads.}
\label{fig:ftbfPads}
\end{figure}
Histograms with responding pads for pions, kaons, and protons can be seen in Fig.~\ref{fig:ftbfPads} in the upper two graphs and left lower. The lower right graph there shows the expected linear scaling (see Eq.~\ref{eq:numberphotons}) of the photon yield with $\sin^2 \theta_C$. A yield of 12 photons per ring for the pion sample can be concluded from these data. This is in better agreement with 16 expected photons per ring at saturated ring diameter.
\begin{figure}[!ht]
\begin{subfigure}{0.5\textwidth}
\centering
\includegraphics[width=3.5in]{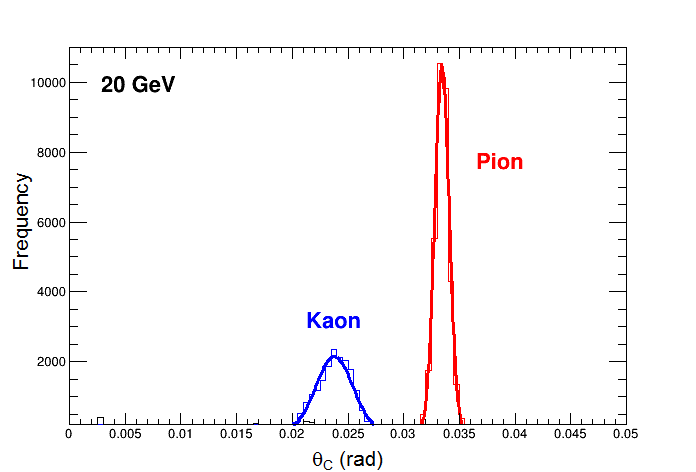}%df}
\end{subfigure}
%\vspace*{-0.5in}
\begin{subfigure}{0.5\textwidth}
\centering
\includegraphics[width=3.5in]{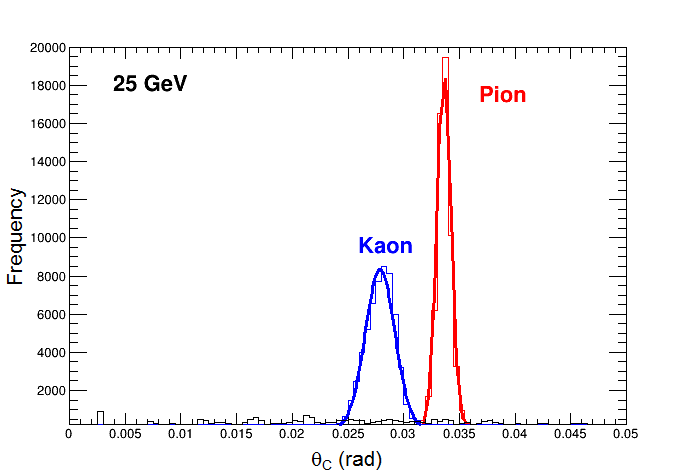}%df}
\end{subfigure}
%\vspace*{-0.5in}
\begin{subfigure}{0.5\textwidth}
\centering
\includegraphics[width=3.5in]{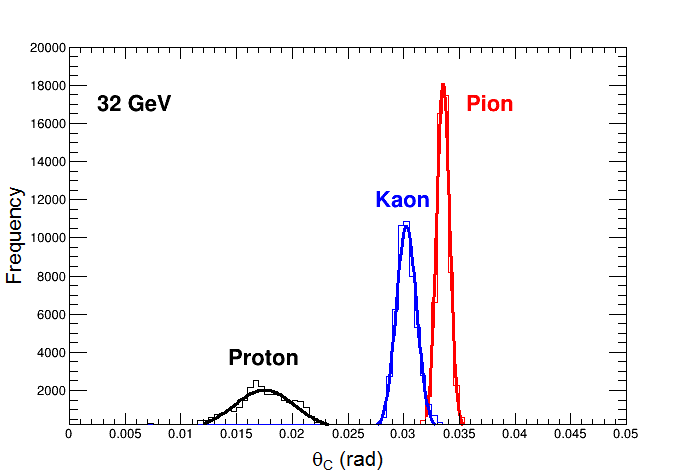}%df}
\end{subfigure}
\caption{(Color online) Particle identification with the RICH detector prototype. Shown are measured Cerenkov angles and Gaussian fits of their distributions, according to the particle species at corresponding energy. Note, that all three graphs show contributions of all three particle species ($\pi$, K, p) and thus "noise" for the proton distribution in the upper two panels where the proton should not appear.}
\label{fig:ftbfRadius}
\end{figure}
The particle identification was performed by comparing the Cerenkov angles measured for known particles at known momenta. As one can see in Fig.~\ref{fig:ftbfRadius} the separation for most abundant particles in a collision experiment is very well achieved.\newline
The peaks in Fig.~\ref{fig:ftbfRadius} can be well understood in terms of angle and width. The angle is determined by the particle velocity. The upper panel of Fig.~\ref{fig:expRadius} shows the correlation between the measurement converting the Cerenkov angle to a ring radius and the calculation with the best fit index of refraction n$_r$=1.00055 (in agreement with published results \cite{ref:cf4index}). The lower panel of the same figure shows the expectation for the width accounting for dispersion in the gas, segmentation of the RICH, momentum spread $\delta p/p=5\%$ of the FNAL beam line (fit as a free parameter), and a constant term of 240 $\mu$m from the fit to account for all other factors. It is assumed that the imperfection of the fit in that panel is influenced by the momentum spread which was assumed to be a constant fractional error, but might vary between different beamline momentum settings.
\begin{figure}[!ht]
\begin{subfigure}{0.5\textwidth}
\centering
\includegraphics[width=3.5in]{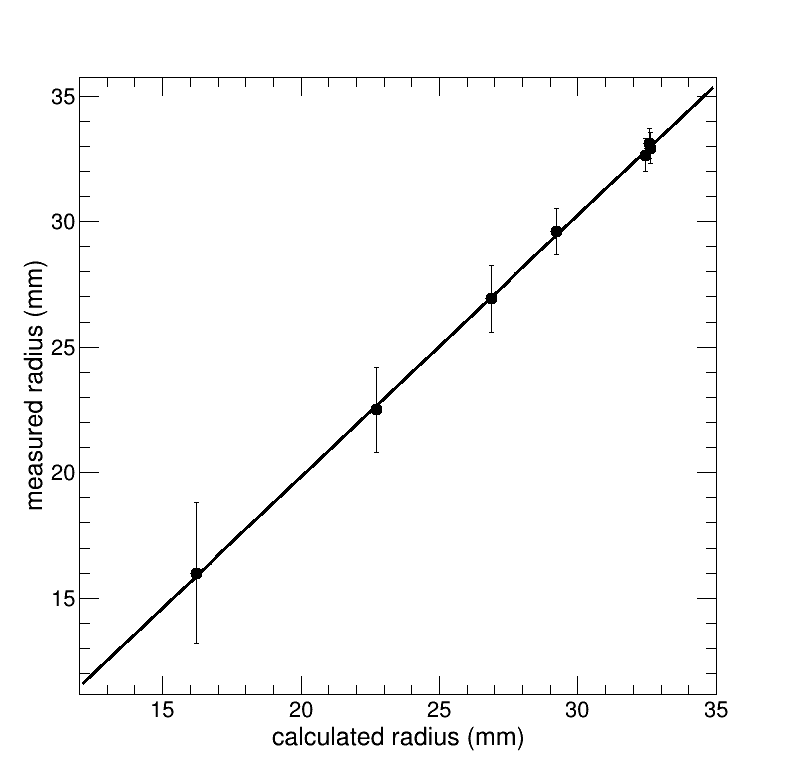}%pdf}
\end{subfigure}
\begin{subfigure}{0.5\textwidth}
\centering
\includegraphics[width=3.5in]{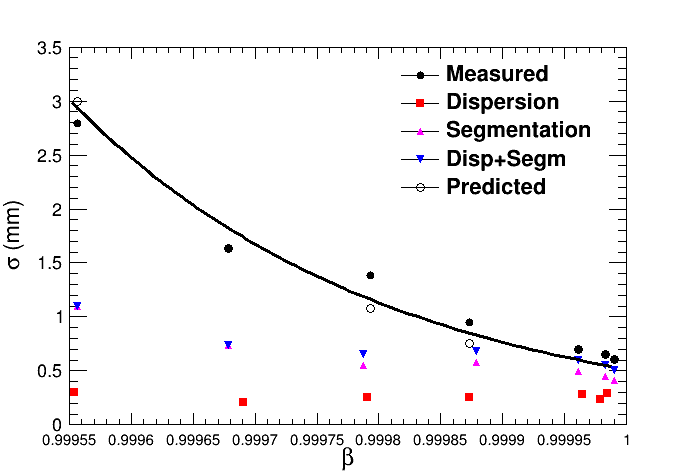}%pdf}
\end{subfigure}
\caption{(Color online) Upper: expected Cerenkov ring radius compared to measured radius based on results in Fig.~\ref{fig:ftbfRadius}. The calculated values were obtained with converting $\cos(\theta_C)=1/(n_r\beta)$ into a radius and varying $n_r$ such that the linear fit gives the best agreement between calculated and measured quantities.
Lower: expected width accounting for various phenomena (see text). Note: the drawn line is visualizing the trend of the measured widths and is not a fitted function.}
\label{fig:expRadius}
\end{figure}
Despite excellent pion-kaon separation out to 32 GeV/c, segmentation of our current prototype (5 mm pads) is nonetheless a significant limiting factor in the ring radius resolution. Because the test beam results are well understood, we can use these results to calculate the pion-kaon separation expectation as a function of photon position resolution (changing only the segmentation term to other pad patterns that result in better position resolution and keeping all other width contributions as measured). Fig.~\ref{fig:piKsep} shows the separation power based on various readout structures that provide a certain position resolution of pads ($\sigma=500~\mu$m,$~300~\mu $m,$~150~\mu$m respectively) for the photons of the Cerenkov ring. The worst assumptions that went into the lower panel of Fig.~\ref{fig:expRadius} were applied to the calculation of the points in Fig.~\ref{fig:piKsep}. If one considers 2.5$\sigma$ separation, which seems to be sufficient because of the expected sample-purity of identified hadrons for the EIC physics program, for discriminating pions from kaons one can see in that figure that the hexagonal readout pads used for this detector prototype will provide separation for momenta up to 45 GeV/c, but for 500 $\mu$m resolution separation can be achieved up to $p\approx$ 60 GeV/c, more than required for the EIC physics program. One can improve the position resolution to 500 $\mu$m and better by achieving charge sharing on readout structures. Further improvements in photon position resolution are not useful since the detector then becomes limited by dispersion in the gas.
\begin{figure}[!ht]
\centering
\includegraphics[width=3.75in]{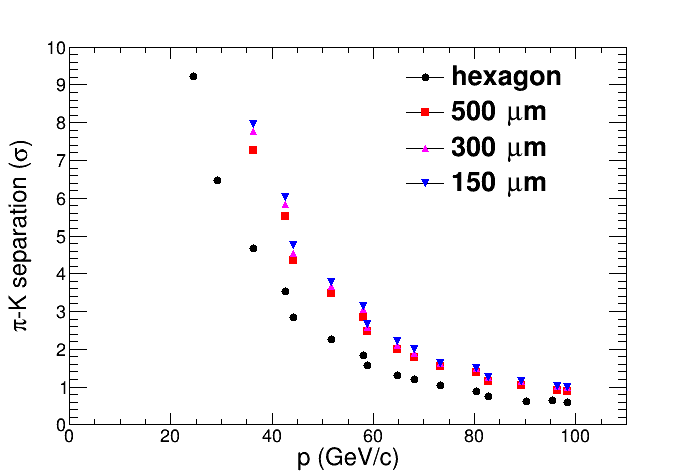}
\caption{(Color online) Pion and kaon separation power: the $\pi$-K separation is defined as the distance from peak to peak in the $\pi$- respectively K-distributions, taken from Fig.~\ref{fig:ftbfRadius} divided by the average of the widths of their distributions. The separation power is increasing when going from hexagons used in this setup to pads with increasing position resolution.}
\label{fig:piKsep}
\end{figure}
%----------------------------------------
%----------------------------------------
\section{Conclusion}
A RICH detector prototype with unique technologies, quintuple-GEM readout structure and dielectric mirror was successfully tested and operated at two test-beam facilities. The first test beam campaign confirmed the proof-of-principle test and validated a set of basic observations as expected which was the measurement of a Cerenkov ring with sufficient number of photons and with a well determined Cerenkov angle, despite the fact that the number of photons was lower than calculated. The second test beam campaign provided the separation of various common particle species in collider experiments ($\pi$, K, p) up to momenta of 32 GeV/c. Very promising results have been obtained, to name one, the rather short radiator size of a RICH but still capable to separate various particle species in the forward direction of collider experiments up to high momenta. One important result is the observation of required position resolution for photon detection to maintain a good separation power in real experimental conditions. The verification of that statement is the subject of further R\&D into an improved readout structure for the RICH-detector and subsequent test in beam.

%----------------------------------------
%----------------------------------------
%End of Document
% or

%%%%%-----BEGIN COMMENT BLOCK
\appendices
\begin{comment}
\section{Proof of the First Zonklar Equation}
Appendix one text goes here.

\section{}
Appendix two text goes here.

\section*{Acknowledgment}
The authors would like to thank the teams of the ESTB, the FTBF, the Stony Brook workshop, and the FLYSUB consortium, who provided us with invaluable help 
\end{comment}
%%%%%-----END COMMENT BLOCK

%\ifCLASSOPTIONcaptionsoff
%  \newpage
%\fi
%\newpage

\IEEEtriggeratref{20}

\begin{thebibliography}{1}
\bibitem{ref:eicwhitepaper}
A.~Accardi et al., "Electron Ion Collider: The Next QCD Frontier - Understanding the glue that binds us all", \href{http://arxiv.org/abs/1212.1701v3}{arXiv:1212.1701v3} (2014)
\bibitem{ref:townmeeting}{\href{https://phys.cst.temple.edu/qcd/}{{2014 Long-range plan, Joint Town Meetings on QCD, Temple University}}}
\bibitem{ref:cf4index} R.~Abjean, A.~Bideau-Menu, and Y.~Quern, "Refractive index of carbon tetrafluoride (CF$_4$) in the 300-140 nm wavelength range", Nucl.~Instrum.~Methods Phys.~Res.~A 292, Issue 3 (1990), 593~-~594
%\bibitem{ref:cf4azmoun}
%B.~Azmoun et al., A Measurement of the Scintillation Light Yield in CF$_4$ Using a Photosensitve GEM Detector, IEEE Transactions on Nuclear Science, Vol.~57, No.~4, August 2010.
\bibitem{ref:cherenkovtamm1}
%\bibitem{Frank:1937fk} 
I.~M.~Frank and I.~Tamm, "Coherent visible radiation of fast electrons passing through matter",
C.\ R.\ Acad.\ Sci.\ URSS {\bf 14}, 109 (1937).
%I.~E.~Tamm, Radiation emitted by uniformly moving electrons, J. Phys.(USSR) 1 (1939) 439.
\bibitem{ref:cherenkovtamm2}
K.~Nakamura et al., Review of particle physics, Journal of Physics G: Nuclear and Particle Physics 37 (2010) 075021
%I.~E.~Tamm, General Characteristics of Vavilov-Cherenkov Radiation, Science 131 (1960), 206–210.
\bibitem{ref:sauligem}
F.~Sauli, GEM: "A new concept for electron amplification in gas detectors", Nucl.~Instrum.~Methods Phys.~Res.~A 386 (1997) 531-534.
\bibitem{ref:hbd4phenix}
W.~Anderson et al., "Design, construction, operation and performance of a Hadron Blind Detector for the PHENIX experiment", Nucl.~Instrum.~Methods Phys.~Res.~A 646 (2011) 35-584
\bibitem{ref:lhcbaging} M. Alfonsi et al., "Aging measurements on triple-GEM detectors operated with CF$_4$-based
gas mixtures", Nuclear Physics B (Proc. Suppl.) 150 (2006) 159–163
\bibitem{ref:4gemgain} D.~M\"ormann, A.~Breskin, R.~Chechik, and D.~Bloch, "Evaluation and reduction of ion back-flow in multi-GEM detectors", Nucl.Instrum.Meth. A516 (2004) 315-326
\bibitem{ref:apv25}
M.~Raymond et al., "The APV25 0.25 $\mu$m CMOS Readout Chip for the CMS Tracker",  Nuclear Science Symposium Conference Record, 2000 IEEE  (Volume:2), 113-118.
\bibitem{ref:srs}
S.~Martiou, H.~Muller, and J.~Toledo, "Front-end electronics for the Scalable Readout System of RD51", Nuclear Science Symposium Conference Record, 2011 IEEE NSS-MIC 2011, N43-5 , 2036-2038.
\bibitem{ref:hbd}
A.~Kozlov, I.~Ravinovich, L.~Shekhtman, Z.~Fraenkel, M.~Inuzuka, and I.~Tserruya, "Development of a triple GEM UV-photon detector operated in pure CF$_4$ for the PHENIX experiment", Nucl.~Instrum.~Methods Phys.~Res.~A 523 (2004) 345-354
\bibitem{ref:phenix}
D.P.~Morrison et al., "The PHENIX experiment at RHIC", Nucl.~Phys.~A638 (1998) 565-570
\bibitem{ref:csicathode}
B.~Azmoun et al., "Collection of Photoelectrons and Operating Parameters of CsI Photocathode GEM Detectors", IEEE Transactions on Nuclear Science, Vol.~56, No.~3, June 2009., 1544 - 1549
\bibitem{ref:purifiers} G.~Baptista, M.~Bosteels, S.~Ilie, C.~Sch\"afer, "Experimental study on oxygen and water removal from gaseous streams for future gas systems in LHC detectors", \url{http://detector-cooling.web.cern.ch/Detector-Cooling/coolingsystems/GasSystemWeb/Document/Gaspurificationreport.PDF}
\bibitem{ref:drs4}
H.~Friederich et al., "A Scalable DAQ System Based on the DRS4 Waveform Digitizing Chip", IEEE Transaction on Nuclear Science, Vol.~ 58, No.~4, August 2011, 1652 - 1656
\bibitem{ref:cht}
D.~Ben-Tzvi and M.~B.~Sandler, "A combinatorial Hough transform", Pattern Recognition
Letters, 11(3):167~-~174, 1990.
\bibitem{ref:sergemmigas}
Jamil A.~Mir et al., "Single-Electron Response Using a GEM-MIGAS Electron Multiplier", IEEE Transactions on Nuclear Science, Vol.~55, No.~4, August 2008
\bibitem{ref:garfield}
R. Veenhof, "Garfield - simulation of gaseous detectors", \url{http://garfield.web.cern.ch/garfield/}, Sept. 1984
\bibitem{ref:magboltz}
S.~Biagi, "Magboltz - transport of electrons in gas mixtures", \url{http://magboltz.web.cern.ch/magboltz/}
\end{thebibliography}
\end{document}